\documentclass[aps,showpacs, nofootinbib]{revtex4}

\input amssym.tex
\input amssym.def

\newcommand{\be}{\begin{equation}}
\newcommand{\ee}{\end{equation}}
\newcommand{\ben}{\begin{eqnarray}}
\newcommand{\een}{\end{eqnarray}}
\def\half{{1 \over 2}}
\newcommand{\la}{{\lambda}}

\newcommand{\cO}{{\cal O}}

\newcommand{\p}{\partial}
\newcommand{\na}{\nabla}

\newcommand{\tchi}{\tilde \chi}

\newcommand{\tim}{{\tilde \mu}}
\newcommand{\tom}{{\tilde \omega}}

\newcommand{\Dsl}{{\slash \negthinspace \negthinspace \negthinspace \negthinspace  D}}

\newcommand{\tV}{{\tilde V}}

\newcommand{\ep}{\epsilon}

\newcommand{\ga}{\gamma}

\pacs{04.50.+h.}

\begin{document}

\title{The Decay of Dirac Hair around a  Dilaton Black Hole}
\author{Gary W.Gibbons}
\affiliation{DAMTP, Centre for Mathematical Sciences,  \protect \\
University of Cambridge\protect \\
Wilberforce Road, Cambridge, CB3 0WA, UK\protect \\
g.w.gibbons@damtp.cam.ac.uk} 
\author{Marek Rogatko}
\affiliation{Institute of Physics \protect \\
Maria Curie-Sklodowska University \protect \\
20-031 Lublin, pl.~Marii Curie-Sklodowskiej 1, Poland \protect \\
rogat@tytan.umcs.lublin.pl \protect \\
rogat@kft.umcs.lublin.pl}

\date{\today}

\begin{abstract}
The intermediate and late-time behaviour of a  massive 
Dirac field in the background of static spherically 
symmetric dilaton black hole solutions is  investigated.
The intermediate asymptotic behaviour of a massive Dirac
field depends  on the mass  parameter  as 
well as the wave number of the  mode,
while the late-time behaviour has 
a  power law decay rate  independent of both.
\end{abstract}

\maketitle

\section{Introduction}

The late-time behaviour of various fields in the spacetime of a collapsing
body is of a great importance in   black hole  physics.
Regardless of details of the collapse or the structure and properties
of the collapsing body the resultant black hole can 
be described by just a few parameters such as
mass, charge and angular momentum, {\it black holes have no hair}. 
The manner and rate with  which the hair of the black hole
decays is thus an important question.  
In what follows we begin by reviewing 
some of the old an new work on this problem.  
\par
Price in Ref.\cite{pri72} for the first time studied the  
neutral external perturbations.  He found 
that the late-time behaviour is 
dominated by the factor $t^{-(2l + 3)}$, for each 
multipole moment $l$. 
The decay along null infinity and along the
future event horizon the fall off was found  \cite{gun94}  
to be like that  $u^{-(l + 2)}$ and $v^{-(l + 3)}$, where $u$ and $v$ were
the outgoing Eddington-Finkelstein and ingoing
Eddington-Finkelstein  coordinates.
In Ref.\cite{bic72} the scalar perturbations on the  Reissner-Nordstr\"om 
background for the case when  $\mid Q \mid < M$ 
was studied. A late time   dependence like  $t^{-(2l + 2)}$
was found,
while for $\mid Q \mid = M$ the late-time behaviour at fixed $r$ is governed by
$t^{-(l + 2)}$. Charged scalar hair 
decayed slower than a neutral one \cite{pir1}-\cite{pir3}, while
the late-time tails in gravitational collapse of 
fields in the background of Schwarzschild solution was reported by 
Burko \cite{bur97}.
The  intermediate and late-time pattern of hair decay was also considered
in Reissner-Nordstr\"om  background in Ref.\cite{hod98}. 
The very late-time tails of  massive scalar fields in the 
Schwarzschild and nearly extremal Reissner-Nordstr/''om black holes
were the subject of \cite{ja}, \cite{ja1}. It was found  that 
the oscillatory
tail of a scalar field decays  like $t^{-5/6}$ at  late times.
Power-law tails in the evolution 
of a charged massless scalar field around the  fixed
background of a  dilaton black hole were 
studied in \cite{mod01a}, while the case of a massive 
scalar field was treated numerically in Ref.\cite{mod01b}.
The analytical proof of the decay pattern both for intermediate and late-time
behaviour was presented in \cite{rog07}. In Ref.\cite{bur04}
the late-time tails of massive scalar field were studied in the spacetime of
stationary axisymmetric black hole and it was found that the power law index of
${-5/6}$ depended neither on multiple mode $l$ nor on the spin rate of the considered black hole.
All the above cases involved bosonic fields.
\par
On the fermionic side, the problem of the late-time behaviour of massive 
Dirac fields were studied
in the spacetime of Schwarzschild black hole \cite{jin04}, 
while in the spacetime of
an Reisner-Nordstr\"om  black hole was analyzed in \cite{jin05}. 
The case of the intermediate and the asymptotic behaviour of charged massive Dirac fields
in the background of Kerr-Newman black hole was elaborated in Ref.\cite{xhe06}.
\par
Interest in 
unification schemes such as superstring/M-theory
has  triggered an  interests in the decay of the hair 
of  $n$-dimensional black holes.
As far as the  $n$-dimensional static black holes is concerned, the
{\it no-hair} theorem for them
is quite well established \cite{unn}. 
The  decay mechanism for  massless scalar hair around
higher $n$-dimensional Schwarzshild spacetime 
was worked out   in \cite{car03}. 
The late-time tails of massive scalar fields in the spacetime of an  $n$-dimensional 
static charged black hole was treated  in \cite{mod05} and it was revealed that
the intermediate asymptotic behaviour of the considered field had the form
$t^{- (l + n/2 - 1/2)}$. The above pattern of decay was confirmed numerically for the case of
$n = 5$ and $n = 6$.
One should also mention the results of Ref.\cite{cho07a}, where the authors obtained 
fermion quasi normal modes for massless Dirac fermion in the background of higher dimensional Schwarzschild black hole.
Recently, there has been also some efforts to study the late-time behaviour of massive scalar fields
in the background of black holes on brane \cite{rog07br} having in mind the idea that our universe is only a submanilfold
on which the standard model is confined to, inside a higher dimensional spacetime \cite{ran99}.
On the other hand, the massless fermion excitations on a tensional 3-brane embedded in six-dimensional spacetime were studied in
\cite{cho07b}.
\par
The main purpose of the present  paper is to extend our knowledge
of the behaviour of fermionic fields and  to 
clarify what kind of mass-induced behaviours play the dominant role
in the asymptotic late-time tails 
as a result of decaying the massive Dirac hair in 
the background of a four-dimensional dilaton black hole. That is in a
spherically symmetric  solution of the low-energy string theory with 
arbitrary coupling constant $\alpha$.
\par
The paper is organized as follows. In Sec.II we gave some general remarks concerning behaviour
of Dirac spinors in a curved background. Sec.III will be devoted to the analytical studies of
intermediate and late-time pattern of decay of the hair in question, while in Sec.IV
we conclude our investigations.

\section{The Dirac Equation in a curved spacetime}

Because of the full spectrum of neutrinos, and their
masses and mixing properties is not known, it seems worth while summarizing
the general situation. Especially as there has recently been some controversy 
\cite{Singh1,Singh2,Pal,Singh3} about the 
so-called Majorana and Dirac masses and their 
consequences  in gravitational field.
\par
In four spacetime dimensions we can always use a representation
in which the gamma matrices are real and we can take the components
of all classical fermion fields
to take values in  a Grassmann algebra over the reals.
\par
The most general Lagrangian for $k$ four-component 
Majorana  fermion $\psi ^i$,
$i=1,2,\dots k$  is 
thus
\be
{\bar \psi}^i \Dsl T_{ij} \psi ^j - {\bar \psi}^i M_{ij} \psi ^j\,,
\ee 
where  $ \Dsl = \gamma ^\mu \na_{\mu}$ and $\na_{\mu}   $ 
 is the covariant derivative 
$\na_{\mu} = \p_{\mu} +
 {1\over 4} \omega_{\mu}^{ab} \ga_{a} \ga_{b}$, $\mu$
and $a$ are tangent and spacetime indices. 
There are related by $e_{\mu}^{a}$, a basis 
of orthonormal one-forms. 
The quantity $\omega_{\mu}^{ab} \equiv \omega^{ab}$ 
are the associated connection
one-forms satisfying
$de^{a} + \omega_{b}{}{}^{a} \wedge e^{b} = 0$. 
On the other hand, $\ga^{\mu}$ are Dirac matrices 
satisfying  $\{ \ga^{a}, \ga^{b} \} =  2 \eta^{ab}$.
We are using a mainly plus metric signature convention. In four spacetime dimensions, for example,  
the gamma matrices may be taken to be real.
The $4k \times 4k$ 
matrices  $T=\tau_{ij}+ \gamma_5 \sigma_{ij}$, 
$M_{ij}= \mu_{ij} +\gamma _5 \nu_{ij}$ with
 $\gamma _5= \gamma_0 \gamma_1 \gamma_2 \gamma _3 $, 
$\gamma_5^2 =-1$ and $ \tau_{ij} , \mu_{ij}, \nu_{ij}
 $ symmetric and $\sigma_{ij}$   anti-symmetric matrices 
which we take  to be independent of time and  position. 
\par
There is an action on  the spinors by  $GL(k,{\Bbb C})$ which satisfies 
\be
\psi^i \rightarrow S^i\,_j \psi ^k,
\ee
where $ S^i\,_j = \exp (\alpha^i\,_j +\gamma _5 \beta ^i\,_j )$
and we are thinking of $\gamma _5 $ as a complex structure on 
${\Bbb  R}^{4k} \equiv {\Bbb C}^{2k}$. Elements of  $ {\Bbb C}^{2k}$
are Weyl (or chiral) spinors for which $\gamma_5 =i$.
Under this action the result yields
\be
T \rightarrow S^\dagger T S\,,\qquad M\rightarrow S^t M S\,. \label{shift}  
\ee 

Using the freedom (\ref{shift}) 
 one may set $T={\rm diag} \pm 1,\pm 1,\dots \pm$.
So as to have positive energy we demand that
all signs are positive.  Now $S \in U(k,{\Bbb C})$ 
and we may choose it to make $M$ diagonal with real non-negative entries
\cite{Zumino}. Thus we arrive at $k$ uncoupled Dirac equations of the form
\be
\bigg( \ga^{\mu} \na_{\mu} - m \bigg) \psi = 0.
\label{dirac}
\ee
If one iterates the Dirac equation and uses the cyclic Bianchi
identity in a curved space one  gets the following:
\be
-\nabla ^2 \psi + {1\over 4} R \psi +  
m^2  \psi  =0.
\ee
There is no  {\it gyro-magnetic} coupling between the
spin and the Ricci or Riemann tensors  \cite{Peres}. 
We see in these calculations no sign of the effect claimed 
in Refs.\cite{Singh1,Singh2,Singh3}.
This is consistent with the {\it equivalence principle}, according to which
all particles should fall in the same way in a gravitational
field. Of course, if the matrices $T$ and $M$ were depended upon position
then, things could be different.
In the presence of a dilaton and axion field, this might  happen.
In this paper we shall just consider mass terms.
Although the detailed calculations above assume that spacetime is
four-dimensional, they are  readily extended to higher dimensional 
spacetimes.

\section{The Decay of Dirac Hair in the  Background of 
a Black Hole Solution}

The treatment of fermions  in spherically symmetric backgrounds
may be greatly simplified by recalling a 
few basic properties of the Dirac 
equation. These allow a rapid reduction of the problem
to the behaviour of a suitable second order radial equation.   
We shall begin by giving  a discussion valid for all spacetime dimensions
$n$ (when $SO(3)$ is replaced by $SO(n-1)$) but our detailed decay
results will apply only to the case $n=4$.

\subsection{Some useful properties of the Dirac operator}
As we saw above,  we may assume that the massive  Dirac equation 
in a  background metric  is given by Eq.(\ref{dirac}).
The basic properties of the Dirac operator 
$\Dsl = \gamma ^\mu \nabla _\mu $ on an $n$-dimensional
manifold that we shall need are
\begin{itemize}
\item for  a metric product
\be
g_{\mu \nu} d x ^\mu d x ^\nu = 
 g_{ab} (x) dx^a dx ^b + g_{mn}(y) dy^m dy ^n\,, 
\ee
it decomposes as a direct sum 
\be
\Dsl = \Dsl_x + \Dsl_y\,.
\ee
\item Under a Weyl conformal  rescaling given by 
\be
g_{\mu \nu} = \Omega ^2 {\tilde g} _{\mu \nu},   
\ee 
it follows directly that we have
\be
\Dsl \psi =\Omega ^{- {1 \over 2} (n+1)} { \tilde {\Dsl} } {\tilde \psi}\,,  
\qquad \psi = \Omega ^ {- {1 \over 2}  (n-1) }\tilde \psi \,.
\ee

\end{itemize}

For a conformo-static metric of the form
\be
ds^2 =-A^2 dt^2 +\Phi ^2 d x ^i d x^i\,,
\ee
where $A=A(x^i)$ and $\Phi=\Phi(x^i)$, $i=1,2\dots n-1$, 
we write
\be
ds^2 =A^2 \biggl( - dt^2 +\bigl ( {\Phi \over A} \bigr) ^2 d x ^i d x^i 
\biggr)\,,
\ee
and consequently find the following:
\be
\Dsl \psi = A^{-{1 \over 2} (n+2)}  \biggl
( \gamma ^0 \partial_t  +\tilde \Dsl  \biggr) \tilde \psi,
\ee
where $\tilde \Dsl$ is the Dirac operator of the metric
$ \bigl ( {\Phi \over A} \bigr) ^2 d x ^i d x^j$ and 
$\tilde \psi= A^{{1 \over 2} (n-1) } \psi$.
Now we use the conformal property again. One obtains the relation 
\be
\tilde \Dsl \tilde \psi = \bigg( {A  \over \Phi }\bigg) ^{ {1 \over 2} (n-1) }
 \gamma ^i \partial_i
\tilde{\tilde \psi}\,, 
\ee 
with $\tilde \psi =  \bigg( {A  \over \Phi } \bigg ) ^{{1\over 2}(n-2)} 
\tilde {\tilde \psi}$.

Since every spherically symmetric metric is conformally flat,
a special case of the theory above, is a static metric of the form given by
\be
ds^2 = -A^2 dt^2 + B^2 dr ^2 + C^2 d \Sigma ^2 _{n-2},
\ee
where $A=A(r), B=B(r), C=C(r)$ are functions only of the radial variable
$r$, and the {\it transverse} metric $d \Sigma ^2 _{n-2} $  
depends neither on $t$ nor on  $r$. 
\par
Let us suppose now, that $\Psi$ is a spinor eigenfunction on the 
$(n-2)$-dimensional  {\it transverse} manifold 
$\Sigma$. Namely, $\Psi$ satisfies the relation of the form as 
\be
\Dsl_\Sigma \Psi = \lambda \Psi.
\ee
By virtue of the properties given above one  
may assume that the following is satisfied:
\be
\Dsl \psi = m \psi, 
\label{mm}
\ee 
and set what follows:
\be
\psi = {1 \over A^{1 \over 2} } { 1 \over C^{(n-2) \over 2 }} \chi \otimes \Psi.
\ee
It can be verified by the direct calculations that Eq.(\ref{mm}) provides the result as
\be
(\gamma ^0 \partial _t + \gamma ^1 \partial _y) \chi = 
A (m- {\lambda \over C} ) \chi.  
\ee
where we have denoted
\be
dy = {B \over A} dr,
\ee
the {\it radial optical distance} (i.e., the Regge-Wheeler radial coordinate).
Gamma matrices $\gamma ^0, \gamma ^1$ 
satisfy the Clifford algebra in two spacetime dimensions.

An identical result may be obtained if a 
Yang-Mills  gauge field $A_{\mu}$ is present 
on the transverse manifold $\Sigma$, but now one gets   
\be
\Dsl_{\Sigma, A_{\mu}}  \Psi = \lambda \Psi,
\ee
where $ \Dsl_{\Sigma, A_{\mu}} $ is 
the Dirac operator twisted by the the connection $A_{\mu}$.

Assuming that $\psi \propto e^{-i\omega t}$, one obtains the
second order equation for $\chi$
\be
{d^2 \chi \over d y^2 }+ \omega ^2 \chi  = 
A^2 (m-{\lambda \over C} )^2 \chi. 
\label{second}
\ee 

In what follows,  the detailed form of the spinor harmonics
and the eigenvalues will not be important. 
\subsection{The Background}

In four spacetime dimensions,  the action for the dilaton gravity 
with arbitrary coupling constant implies
\be
S = \int d^4x \sqrt{-g} 
\bigg[ R - 2 \na^{\mu} \phi \na_{\mu} \phi - e^{- 2 \alpha \phi} 
F_{\mu \nu}  F^{\mu \nu} \bigg],
\ee
where $\phi$ is the dilaton field, $\alpha$ 
coupling constant while $F_{\mu \nu} = 2 \na_{[\mu} A_{\nu]}$
is the strength of $U(1)$ gauge field.\\
The static spherically symmetric solution of the equations  of motion 
are given by the following line element:
\be
ds^2 = - \bigg( 1 - {r_{+} \over r}
 \bigg)\bigg( 1 - {r_{-} \over r}
 \bigg)^{{1 - \alpha^2} \over 1 + \alpha^2} dt^2
+ {dr^2 \over \bigg( 1 - {r_{+} \over r} \bigg)
\bigg( 1 - {r_{-} \over r} \bigg)^{{1 - \alpha^2} \over 1 + \alpha^2}}
+ R^2(r) d\Omega^2,
\label{dila}
\ee
where $R^2(r) = r^2 \bigg( 1 - {r_{-} \over r} \bigg)^{2 \alpha^2 \over 1 + \alpha^2}$,
 while $r_{+}$ and $r_{-}$
are related to the mass $M$ and the electric 
charge $Q$ of the black hole
\be
e^{- 2 \alpha \phi} = \bigg( 1 - {r_{-} \over r} \bigg)^{2 \alpha^2 \over 1 + \alpha^2}, \qquad
2 M = r_{+} + {1 - \alpha^2 \over 1 + \alpha^2} r_{-}, \qquad
Q^2 = {r_{-}~r_{+} \over 1 + \alpha^2}.
\ee
The metric is asymptotically flat in the sense that
the spacetime  
contains an initial  data set
$(\Sigma_{end}, g_{ij}, K_{ij})$ with gauge fields such that 
$\Sigma_{end}$ is diffeomorphic to ${\bf R}^3$ minus a ball and the 
following asymptotic conditions are fulfilled:
\ben
\vert g_{ij}  - \delta_{ij} \vert + r \vert \p_{a}g_{ij} \vert
+ ... + r^k \vert \p_{a_{1}...a_{k}}g_{ij} \vert +
r \vert K_{ij} \vert + ... + r^k \vert \p_{a_{1}...a_{k}}K_{ij} \vert
\le {\cal O}\bigg( {1\over r} \bigg), \\
\vert F_{\alpha \beta} \vert + r \vert \p_{a} F_{\alpha \beta} \vert
+ ... + r^k \vert \p_{a_{1}...a_{k}}F_{\alpha \beta} \vert
\le {\cal O}\bigg( {1 \over r^2} \bigg),\\
\phi = \phi_{0} + {\cal O}\bigg( {1\over r} \bigg),
\een
where $K_{ij}$ is the exterior curvature, $\phi_{0}$
 is a constant value of the scalar field.\\

\subsection{Spinor No-Hair theorems}

The properties of static spinor fields around Schwarzschild and Kerr
black holes
and the consequent {\it no-hair} properties
have been investigated by many people including 
\cite{Hartle1,Hartle2,Teitelboim1,Teitelboim2}.   
The basic idea is to study solutions of the static Dirac
on the background of the black hole. One either considers the case when
there are no fermionic sources outside the horizon or
one constructs a Green  function. 
In the massless static spherical case it is clear from
our work above that this is equivalent to solving the flat space
Dirac equation \cite{Gibbons82}
\be
\gamma ^i \partial _i  \tilde {\tilde \psi }=0.
\ee  
This may have regular solutions on the horizon. On the other hand,
in our case, we have the following:
\be
\psi = { 1 \over A \Phi ^{ \half (n-2) } }   \tilde {\tilde \psi} \,.
\ee
On the horizon $ A=0$ and, unless the solution is extreme, $\Phi \ne 0$
{\cite{Gibbons82}.
The extreme case is exceptional because ${ 1 \over \Phi}  = 0$ at 
the horizon in such a way that the spinor $\psi$ remains finite
\cite{Gibbons82}.
Of course, in the non-extreme case, 
one should check that some scalar spinorial invariant
blows up, but this can easily be done.  
 
\subsection{Decay of Fermionic Hair}
We shall now analyze the time evolution of massive Dirac spinor field in 
the background of
dilaton black hole 
by means of the spectral decomposition method.
In \cite{hod98},\cite{lea86} it was 
shown that the asymptotic tail is connected with the
existence of a branch cut situated along the interval $-m \le \omega \le m$.
An oscillatory inverse power-law behaviour of massive Dirac field arises
from the integral of Green function $\tilde G(y, y';\omega)$ around branch cut.
The time evolution of massive Dirac field may be written in the following form:
\be
\chi(y, t) = \int dy' \bigg[ G(y, y';t) \chi_{t}(y', 0) +
G_{t}(y, y';t) \chi(y', 0) \bigg],
\ee
for $t > 0$, where   the Green's function  $ G(y, y';t)$ is given by the relation
\be
\bigg[ {\p^2 \over \p t^2} - {\p^2 \over \p y^2 } + V \bigg]
G(y, y';t)
= \delta(t) \delta(y - y').
\label{green}
\ee
In what follows, 
our main task will be to find the dilaton black hole Green function.
Using the Fourier transform \cite{lea86}
$\tilde  
G(y, y';\omega) = \int_{0^{-}}^{\infty} dt~ G(y, y';t) e^{i \omega t}$ one can
reduce equation
(\ref{green}) to an ordinary differential equation.
The Fourier's transform is well defined for $Im~ \omega \ge 0$, while the 
corresponding inverse transform yields
\be
G(y, y';t) = {1 \over 2 \pi} \int_{- \infty + i \ep}^{\infty + i \ep}
d \omega~
\tilde G(y, y';\omega) e^{- i \omega t},
\ee
for some positive number $\ep$.
By virtue of the above
the Fourier's component of the Green's function $\tilde  G(y, y';\omega)$
can be written in terms of two linearly independent solutions for
homogeneous equation. Namely, one has
\be
\bigg(
{d^2 \over dy^2} + \omega^2 - \tV \bigg) \chi_{i} = 0, \qquad i = 1, 2,
\label{wav}
\ee
where $\tV = A^{2} \bigg( m - {\la \over C} \bigg)^2$.\\

The boundary conditions for $\chi_{i}$ are described by purely ingoing waves
crossing the outer horizon $H_{+}$ of the 
static charged black hole
$\chi_{1} \simeq e^{- i \omega y}$ as $y \rightarrow  - \infty$. On the other hand, 
$\chi_{2}$ should be damped exponentially at $i_{+}$, namely
$\chi_{2} \simeq e^{- \sqrt{m^2 - \omega^2}y}$ at $y \rightarrow \infty$.
\par
Let us assume that the observer and the initial data are situated far away from the considered
black hole. In order to rewrite Eq.(\ref{wav}) in a more convenient form we change variables
\be
\chi_{i} = {\xi \over \bigg( 1 - {r_{+} \over r} \bigg)^{1/2}
\bigg( 1 - {r_{-} \over r} \bigg)^{{1 - \alpha^2} \over 2(1 + \alpha^2)}},
\ee
where $i = 1,2$. 
Then, one can 
expand Eq.(\ref{wav}) as a power  series of $ r_{\pm}/r$ neglecting terms of order
$\cO ((\omega/r)^2)$ and higher. Under this assumption we reach to the following:
\ben \label{whit}
{d^2 \over dr^2} \xi &+& \bigg[
\omega^2 - m^2 + {2 \omega^2 ( r_{+} + \alpha_{1} r_{-}) 
- m^2 ( r_{+} + \alpha_{1} r_{-})
+ 2 \la m (1 + r_{+})
\over r} \\ \nonumber
&-& {\la^2 - 2 \la m r_{-} ( \alpha_{1} + \alpha_{2}) + m^2 \alpha_{1} r_{+} r_{-}
\over r^2}
\bigg] \xi = 0,
\een
where we have denoted $\alpha_{1} = {1 - \alpha^2 \over 1 + \alpha^2}$ and 
$\alpha_{2} = {2\alpha^2 \over 1 + \alpha^2}$.\\
It can be verified that Eq.(\ref{whit}) may be solved in terms of Whittaker's functions. 
Consequently, two basic solutions are needed to construct the Green function, with the condition that
$\mid \omega \mid \ge m$. 
Namely, the Whittaker's functions 
$\tchi_{1} = M_{\delta, \tim}(2 \tom r)$ and $\tchi_{2} = W_{\delta, \tim}(2 \tom r)$
have the following parameters:
\ben
\tim &=& \sqrt{ 1/4 + \la^2 - 2 \la m r_{-} + m^2 \alpha_{1} r_{+} r_{-} }, \\ \nonumber
\delta &=&  {\omega^2 ( r_{+} + \alpha_{1} r_{-}) + \la m (1 + r_{+}) - {m^2 \over 2}( r_{+} + \alpha_{1} r_{-})
\over \tom},\\ \nonumber
\tom^2 &=& m^2 - \omega^2.
\een
On the other hand, the spectral Green function takes the form as
\ben
G_{c}(r, r';t) &=& {1 \over 2 \pi} \int_{-m}^{m}dw
\bigg[ {\tchi_{1}(r, \tom e^{\pi i})~\tchi_{2}(r',\tom e^{\pi i}) \over W(\tom e^{\pi i})}
- {\tchi_{1}(r, \tom )~\tchi_{2}(r',\tom ) \over W(\tom )} 
\bigg] ~e^{-i w t} \\ \nonumber
&=& {1 \over 2 \pi} \int_{-m}^{m} dw f(\tom)~e^{-i w t},
\een 
where $W(\tom)$ is the Wronskian.\\
Further, we focus our attention on
the intermediate asymptotic decay of the massive Dirac hair, i.e., in the range of parameters
$M \ll  r \ll t \ll M/(m M)^2$.
The intermediate asymptotic contribution to the Green function integral gives the frequency equal to 
$\tom = {\cO (\sqrt{m/t})}$, which in turns implies that $\delta \ll 1$. Having in mind that $\delta$ 
results from the $1/r$ term in the massive scalar field equation of motion, it depicts
the effect of backscattering off the spacetime curvature and in the case under consideration
the backscattering is negligible. Taking into account all the above and the fact that $\tom r \ll 1$ 
and $M(a, b, z) = 1$ as $z$ tends to zero, we obtain the resulting expression for spectral Green function
\be
G_{c}(r, r';t) = {2^{3 \tim - {3\over2}} \over \tim \sqrt{\pi}}
{\Gamma(-2\tim)~\Gamma({1 \over 2} + \tim) \Gamma(\tim +1 ) \over
\tim \Gamma(2 \tim)~\Gamma({1 \over 2} - \tim)}
\bigg( 1 + e^{(2 \tim + 1) \pi i} \bigg)~(r r')^{{1 \over 2} + \tim} 
~\bigg( {m\over t} \bigg)^{{1 \over 2} + \tim}~J_{{1 \over 2} + \tim}(mt).
\ee  
In the limit when $t \gg 1/m$ it implies
\be
G_{c}(r, r';t) = {2^{3 \tim - 1} \over \tim \sqrt{\pi}}
{\Gamma(-2\tim)~\Gamma({1 \over 2} + \tim) \Gamma(\tim +1 ) \over
\tim \Gamma(2 \tim)~\Gamma({1 \over 2} - \tim)}
\bigg( 1 + e^{(2 \tim + 1) \pi i} \bigg)~(r r')^{{1 \over 2} + \tim} 
~m^{\tim}~ t^{- 1 - \tim}~\cos(mt - {\pi \over 2}(\tim + 1)).
\label{gfim}
\ee  
Eq.(\ref{gfim}) depicts the oscillatory inverse power-law behaviour. We remark that in our case the intermediate
times of the power-law tail depends only on $\tim$ which in turn is a function of the multiple number
of the wave modes.
\par
The different pattern of decay is expected when  $\kappa \gg 1$, for the late-time
behaviour, when the backscattering off the curvature is important.
Consequently, $f(\tom)$ when $\kappa \gg 1$ may be rewritten in the following form:
\ben \label{fer}
f(\tom) &=& {\Gamma(1 + 2\tim)~\Gamma(1 - 2\tim) \over 2 \tim}~(r r')^{1 \over 2}
\bigg[ J_{2 \tim} (\sqrt{8 \delta \tom r})~J_{- 2 \tim} (\sqrt{8 \delta \tom r'})
- I_{2 \tim} (\sqrt{8 \delta \tom r})~I_{- 2 \tim} (\sqrt{8 \delta \tom r'}) \bigg] \\ \nonumber
&+&
{(\Gamma(1 + 2\tim))^2~\Gamma(- 2\tim)~\Gamma( {1 \over 2} + \tim - \delta)
 \over 2 \tim ~\Gamma(2 \tim)~\Gamma({1 \over 2} - \tim - \delta) }~(r r')^{1 \over 2}
~\delta^{- 2 \tim}
\bigg[
J_{2 \tim} (\sqrt{8 \delta \tom r})~J_{2 \tim} (\sqrt{8 \delta \tom r'})
\\ \nonumber
&+& e^{(2 \tim + 1)}
I_{2 \tim} (\sqrt{8 \delta \tom r})~I_{2 \tim} (\sqrt{8 \delta \tom r'}) 
\bigg],
\een
where we have used the limit $M_{\delta, \tim}(2 \tom r) \approx
\Gamma (1 + 2 \tim) (2 \tom r)^{1 \over 2}~\delta^{- \tim}~J_{\tim}(\sqrt{8 \delta \tom r})$.
The first part of the above Eq.(\ref{fer}) the late time tail is proportional to $t^{-1}$
and it occurs that we shall concentrate on
the second term of the right-hand side of Eq.(\ref{fer}). It turned out that for the case when 
$\kappa \gg 1$ it may be rewritten in the form as
\be
G_{c~(2)}(r, r';t) = {M \over 2 \pi} \int_{-m}^{m}~dw~e^{i (2 \pi \delta - wt)}~e^{i \varphi},
\label{gc2}
\ee
where we have used the following definition:
\be
e^{i \varphi} = { 1 + (-1)^{2 \tim} e^{- 2 \pi i \delta} \over
 1 + (-1)^{2 \tim} e^{2 \pi i \delta}}.
\ee
On the other hand, $M$ yields
\be
M = {(\Gamma(1 + 2\tim))^2~\Gamma(- 2\tim) \over 2 \tim ~\Gamma(2 \tim) }~(r r')^{1 \over 2}
\bigg[
J_{2 \tim} (\sqrt{8 \delta \tom r})~J_{2 \tim} (\sqrt{8 \delta \tom r'})
+ I_{2 \tim} (\sqrt{8 \delta \tom r})~I_{2 \tim} (\sqrt{8 \delta \tom r'}) 
\bigg].
\ee
At very late time both terms $e^{i w t}$ and $e^{2 \pi \delta}$ are rapidly
oscillating. 
From this fact it follows directly that 
the spinor waves are mixed states consisting of the states 
with multipole phases backscattered by spacetime curvature, which most of them cancel
with each others which have the inverse phase. Thus, one can find the value of 
$G_{c~(2)}$ by means of the saddle point method. 
The saddle point integration allows us to evaluate the accurate value of the asymptotic behaviour.
Namely,
it could be found that the value $2 \pi \delta
- wt$ is stationary at the value of $w$ equal to the following:
\be
a_{0} = \bigg[ {\pi~(\omega^2~ (r_{+} + \alpha_{1} r_{-}) +
\la m (1 + \alpha_{1} r_{-})  - {m^2 \over 2} (r_{+} + \alpha_{1} r_{-}) 
\over  \sqrt{2} m} \bigg]^{1 \over 3},
\ee
Evaluating Eq.(\ref{gc2}) by means of the saddle point integration we achieve finally to the form of
the spectral Green function for massive Dirac spinor hair. It implies
\be            
G_{c}(r, r';t) = { 2 \sqrt{2} \over \sqrt{3}}~m^{2/3}~ (\pi)^{5 \over 6}
\bigg[
2 m^2 (r_{+} + \alpha_{1} r_{-}) + 2 \la m (1 + r_{+}) - m^2 (r_{+} + \alpha_{1} r_{-}) 
\bigg]^{1 \over 3}
(mt)^{-{ 5 \over 6}}~\sin(mt)~\tchi(r, m)~\tchi(r', m).
\label{spg}
\ee
The above equation provides the main result of our calculations. 
It illustrates  the fact that
the late-time asymptotic decay pattern of massive Dirac hair in the background of spherically symmetric
dilaton black hole is proportional to $- 5/6$.



\section{Conclusions}
In this  paper we have treated 
the problem of the asymptotic tail behaviour of a free 
Dirac field in the spacetime  of a  spherically symmetric charged black hole 
solution of dilaton
gravity with arbitrary coupling constant $\alpha$. 
This theory is related to the the low-energy
limit of the heterotic string theory and on its own is a generalization of 
electromagnetism by adding scalar field
$\phi$ dilaton and coupling constant between $U(1)$ gauge field and 
scalar field.
\par
The resultant intermediate asymptotic behaviour depends on the 
field parameter mass as well as 
the wave number of the  mode. 
But this is not the final pattern of 
decay of the  massive Dirac hair.
Resonance backscattering off the spacetime curvature 
dominates at late times. 
We have calculated analytically
that the pattern of decay in question is proportional to 
$t^{-{ 5 \over 6}}$. The same result one gets 
studying the late-time behaviour of free massive scalar fields in the 
same  background. One should remark that the above considerations are also applicable to the case
of extremal dilaton black hole, i.e., to the case when $r_{+} = r_{-}$. Thus having in mind Eq.(\ref{spg})
one gets the exact form of the spectral Green function for the late-time behaviour of massive Dirac hair for
the extremal dilaton black hole in the theory with arbitrary coupling constant $\alpha$.

\begin{acknowledgments}
MR is grateful for hospitality of DAMTP, Center for Mathematical Sciences, Cambridge, where 
the part of the research was begun.
This work was partially financed by the Polish budget funds in 2007 year as the research project. 
\end{acknowledgments}





\begin{thebibliography}{99}
%
\def\cmp#1#2#3{{ Commun. Math. Phys.} {\bf #1}, #2 (#3)}
\def\lmp#1#2#3{{ Lett. Math. Phys.} {\bf #1}, #2 (#3)}
\def\hpa#1#2#3{{ Hell. Phys. Acta} {\bf #1}, #2 (#3)}
\def\grg#1#2#3{{ Gen. Rel. Grav.} {\bf #1}, #2 (#3)}
\def\pr#1#2#3{{ Phys. Rev.} {\bf #1}, #2 (#3)}
\def\prl#1#2#3{{ Phys. Rev. Lett.} {\bf #1}, #2 (#3)}
\def\prd#1#2#3{{ Phys. Rev. D} {\bf #1}, #2 (#3)}
\def\pl#1#2#3{{ Phys. Lett} {\bf #1}, #2 (#3)}
\def\pla#1#2#3{{ Phys. Lett. A} {\bf #1}, #2 (#3)}
\def\plb#1#2#3{{ Phys. Lett. B} {\bf #1}, #2 (#3)}
\def\prep#1#2#3{{ Phys. Reports} {\bf #1}, #2 (#3)}
\def\phys#1#2#3{{ Physica} {\bf #1}, #2 (#3)}
\def\jcp#1#2#3{{ J. Comput. Phys.} {\bf #1}, #2 (#3)}
\def\jmp#1#2#3{{ J. Math. Phys.} {\bf #1}, #2 (#3)}
\def\jpm#1#2#3{{ J. Phys. A: Math. Gen.} {\bf #1}, #2 (#3)}
\def\cpr#1#2#3{{ Computer Phys. Rept.} {\bf #1}, #2 (#3)}
\def\cqg#1#2#3{{ Class. Quantum Grav.} {\bf #1}, #2 (#3)}
\def\cma#1#2#3{{ Computers Math. Applic.} {\bf #1}, #2 (#3)}
\def\mc#1#2#3{{ Math. Compt.} {\bf #1}, #2 (#3)}
\def\apj#1#2#3{{ Astrophys. J.} {\bf #1}, #2 (#3)}
\def\apjs#1#2#3{{ Astrophys. J. Suppl.} {\bf #1}, #2 (#3)}
\def\acta#1#2#3{{ Acta Astronomica} {\bf #1}, #2 (#3)}
\def\apl#1#2#3{{Ann. Physik. (Leipzig)} {\bf #1}, #2 (#3)}
\def\anp#1#2#3{{Ann. Phys. } {\bf #1}, #2 (#3)}
\def\sa#1#2#3{{ Sov. Astro.} {\bf #1}, #2 (#3)}
\def\sia#1#2#3{{ SIAM J. Sci. Statist. Comput.} {\bf #1}, #2 (#3)}
\def\aa#1#2#3{{ Astron. Astrophys.} {\bf #1}, #2 (#3)}
\def\mnras#1#2#3{{ Mon. Not. R. astr. Soc.} {\bf #1}, #2 (#3)}
\def\npb#1#2#3{{ Nucl. Phys. B} {\bf #1}, #2 (#3)}
\def\prsla#1#2#3{{ Proc. R. Soc. London, Ser. A} {\bf #1}, #2 (#3)}
\def\jhep#1#2#3{{ JHEP} {\bf #1}, #2 (#3)}
\def\nuc#1#2#3{{Nuovo Cimento B } {\bf #1}, #2 (#3)}
\def\lnuc#1#2#3{{Lett. Nuovo Cimento } {\bf #1}, #2 (#3)}
\def\ijmp#1#2#3{{Int. J. Mod. Phys. D} {\bf #1}, #2 (#3)}
\def\atmp#1#2#3{{Adv. Theor. Math. Phys.} {\bf #1}, #2 (#3)}
\def\ptps#1#2#3{{Prog. Theor. Phys. Suppl.} {\bf #1}, #2 (#3)}
\def\ptp#1#2#3{{Prog. Theor. Phys. } {\bf #1}, #2 (#3)}
\def\lmp#1#2#3{{Lett. Math. Phys. } {\bf #1}, #2 (#3)}
%
\def\hepph#1#2{{ hep-ph }{\bf #1} (#2)}
\def\hepth#1#2{{ hep-th }{\bf #1} (#2)}
\def\grqc#1#2{{ gr-qc }{\bf #1} (#2)}
\def\ibid#1#2#3{{ {\it ibid.} }{\bf #1}, #2 (#3)}
%

\bibitem{pri72}
R.H.Price, \prd{5}{2419}{1972}.
\bibitem{gun94}
C.Gundlach, R.H.Price and J.Pullin, \prd{49}{883}{1994}.
\bibitem{bic72}
J.Bicak, \grg{3}{331}{1972}.
\bibitem{pir1}
S.Hod and T.Piran, \prd{58}{024017}{1998}.
\bibitem{pir2}
S.Hod and T.Piran, \prd{58}{024018}{1998}.
\bibitem{pir3}
S.Hod and T.Piran, \prd{58}{024019}{1998}.
\bibitem{bur97}
L.M.Burko, {\it Abstracts of plenary talks and contributed papers},
15th International Conference on General Relativity and Gravitation,
Pune, 1997, p.143, unpublished.
\bibitem{hod98}
S.Hod and T.Piran, \prd{58}{044018}{1998}.
\bibitem{ja}
H.Koyama and A.Tomimatsu, \prd{63}{064032}{2001}.
\bibitem{ja1}
H.Koyama and A.Tomimatsu, \prd{64}{044014}{2001}.
\bibitem{mod01a}
R.Moderski and M.Rogatko, \prd{63}{084014}{2001}.
\bibitem{mod01b}
R.Moderski and M.Rogatko, \prd{64}{044024}{2001}.
\bibitem{rog07}
M.Rogatko, \prd{75}{10406}{2007}.
\bibitem{bur04}
L.M.Burko and G.Khanna, \prd{70}{044018}{2004}.
\bibitem{jin04}
J.L.Jing, \prd{70}{065004}{2004}.
\bibitem{jin05}
J.L.Jing, \prd{72}{027501}{2005}.
\bibitem{xhe06}
X.He and J.L.Jing, \npb{755}{313}{2006}.

\bibitem{unn}
G.W.Gibbons, D.Ida and T.Shiromizu, \ptps{148}{284}{2003},\\
G.W.Gibbons, D.Ida and T.Shiromizu, \prl{89}{041101}{2002},\\
G.W.Gibbons, D.Ida and T.Shiromizu, \prd{66}{044010}{2002},\\
M.Rogatko, \cqg{19}{L151}{2002},\\
M.Rogatko, \prd{67}{084025}{2003},\\
M.Rogatko, \prd{70}{044023}{2004},\\
M.Rogatko, \prd{71}{024031}{2005},\\
M.Rogatko, \prd{73}{124027}{2006}.
\bibitem{car03}
V.Cardoso, S.Yoshida and O.J.C.Dias, \prd{68}{061503}{2003}.
\bibitem{mod05}
R.Moderski and M.Rogatko, \prd{72}{044027}{2005}.
\bibitem{cho07a}
H.T.Cho, A.S.Cornell, J.Doukas, and W.Naylor, \prd{75}{104005}{2007}.
\bibitem{rog07br}
M.Rogatko and A.Szyplowska, \prd{76}{044010}{2007}.
\bibitem{ran99}
L.Randall and R.Sundrum, \prl{83}{3370}{1999}.
\bibitem{cho07b}
H.T.Cho, A.S.Cornell, J.Doukas, and W.Naylor, {\it Fermion Excitation on a Tense Brane Black Hole},
\hepth{0710.5267}{2007}.

\bibitem{lea86}
E.W.Leaver, \prd{34}{384}{1986}.


\bibitem{Singh1}
  D.~Singh, N.~Mobed and G.~Papini,
{\it The distinction between Dirac and Majorana neutrino wave packets due to
  gravity and its impact on neutrino oscillations}, \grqc{0606134}{2006}.

\bibitem{Singh2}
 D.~Singh, N.~Mobed and G.~Papini, \prl{97}{041101}{2006}.

\bibitem{Pal}
  J.~F.~Nieves and P.~B.~Pal,
{\it  Comment on 'Can gravity distinguish between Dirac and Majorana
 neutrinos?}, \grqc{0610098}{2006}.
  

\bibitem{Singh3}
  D.~Singh, N.~Mobed and G.~Papini,
 {\it `Reply to comment on 'Can gravity distinguish between Dirac and Majorana
  neutrinos?}, \grqc{0611016}{2006}.
 
\bibitem{Zumino}
B.Zumino, \jmp{3}{1055}{1962}.

\bibitem{Akhmedov} 
E.Kh.Akhmedov, {\it Neutrino Physics}, \hepph{0001264}{2000}.


\bibitem{Peres} 
 A.Peres, \nuc{28}{1091}{1963}.


\bibitem{Hartle1}
  J.~B.~Hartle, \prd{3}{2938}{1971}.
 
\bibitem{Hartle2}
  J.~B.~Hartle,
{\it Can A Schwarzschild Black Hole Exert Long Range Neutrino Forces?} in ,
J R Klauder, {\it Magic Without Magic}, 259-275  (San Francisco 1972).    

\bibitem{Teitelboim1}
  C.~Teitelboim, \lnuc{3}{326}{1972}.

\bibitem{Teitelboim2}
  C.~Teitelboim, \lnuc{3}{397}{1972}.

\bibitem{Gibbons82}
  G.~W.~Gibbons,
The Multiplet Structure Of Solitons In The O(2) Supergravity Theory,'
PRINT-82-0183-CAMBRIDGE
published as
The Multiplet Structure of Solitons in the O(2)
Supergravity Theories in {\it  Quantum structure of space and time} eds.
 M J Duff \& C J Isham, 317­321 (Cambridge University Press, Cambridge 1983).


\end{thebibliography}
\end{document}